\documentclass[%
 reprint,
superscriptaddress,
 amsmath,amssymb,
 aps,
 prl,
]{revtex4-1}

\usepackage{graphicx}% Include figure files
\usepackage{dcolumn}% Align table columns on decimal point
\usepackage{bm}% bold math
\usepackage{xcolor}
\usepackage{multirow}
\usepackage{threeparttable}
\usepackage{color}
\usepackage[ulem=normalem]{changes}
\definechangesauthor[name={Yassid Ayyad}, color={red}]{YA}

\begin{document}

\title{ Near-threshold dipole strength in $^{10}$Be with isoscalar character} % Force line breaks with \\

\author{J.~Chen}
\email{chenjie@sustech.edu.cn}
\affiliation{College of Science, Southern University of Science and Technology, Shenzhen, 518055, Guangdong, China}
\affiliation{Physics Division, Argonne National Laboratory, Lemont, Illinois 60439, USA}

\author{Y.~Ayyad}
\email{yassid.ayyad@usc.es}
\affiliation{%
IGFAE, Universidade de Santiago de Compostela, E-15782, Santiago de Compostela, Spain}

\author{D.~Bazin}
\affiliation{%
FRIB/NSCL Laboratory, Michigan State University, East Lansing, Michigan 48824, USA}

\author{W.~Mittig}
\affiliation{%
FRIB/NSCL Laboratory, Michigan State University, East Lansing, Michigan 48824, USA}

\author{M.~Z.~Serikow}
\affiliation{%
FRIB/NSCL Laboratory, Michigan State University, East Lansing, Michigan 48824, USA}

\author{N.~Keeley}
\affiliation{National Centre for Nuclear Research, ul.\ Andrzeja So\l tana 7, 05-400 Otwock, Poland}

\author{S.~M.~Wang}
\affiliation{Key Laboratory of Nuclear Physics and Ion-beam Application (MOE), Institute of Modern Physics, Fudan University, Shanghai 200433, China}
\affiliation{Shanghai Research Center for Theoretical Nuclear Physics,
NSFC and Fudan University, Shanghai 200438, China}

\author{B.~Zhou}
\affiliation{Key Laboratory of Nuclear Physics and Ion-beam Application (MOE), Institute of Modern Physics, Fudan University, Shanghai 200433, China}
\affiliation{Shanghai Research Center for Theoretical Nuclear Physics,
NSFC and Fudan University, Shanghai 200438, China}

\author{J.~C.~Zamora}	
\affiliation{%
FRIB/NSCL Laboratory, Michigan State University, East Lansing, Michigan 48824, USA}

\author{S.~Beceiro-Novo}
\affiliation{%
Universidade da Coruña, Campus Industrial, Departamento de Física y Ciencias de la Tierra, CITENI, Ferrol, 15471, Spain
}

\author{M.~Cortesi}
\affiliation{%
FRIB/NSCL Laboratory, Michigan State University, East Lansing, Michigan 48824, USA}

\author{M.~DeNudt}
\affiliation{%
FRIB/NSCL Laboratory, Michigan State University, East Lansing, Michigan 48824, USA}

\author{S.~Heinitz}
\affiliation{Laboratory of Radiochemistry, Paul Scherrer Institute, Villigen, Switzerland}

\author{S.~Giraud}
\affiliation{%
FRIB/NSCL Laboratory, Michigan State University, East Lansing, Michigan 48824, USA}

\author{P.~Gueye}	
\affiliation{%
FRIB/NSCL Laboratory, Michigan State University, East Lansing, Michigan 48824, USA}

\author{C.~R.~Hoffman}
\affiliation{Physics Division, Argonne National Laboratory, Lemont, Illinois 60439, USA}

\author{B.~P.~Kay}
\affiliation{Physics Division, Argonne National Laboratory, Lemont, Illinois 60439, USA}

\author{E.~A.~Maugeri}	
\affiliation{Laboratory of Radiochemistry, Paul Scherrer Institute, Villigen, Switzerland}

\author{B.~G.~Monteagudo}
\affiliation{%
FRIB/NSCL Laboratory, Michigan State University, East Lansing, Michigan 48824, USA}
\affiliation{%
Department of Physics, Hope College, Holland, Michigan 49422-9000, USA}

\author{H.~Li}
\affiliation{Institute of Modern Physics, Chinese Academy of Sciences, Lanzhou 730000, China}

\author{W.~P.~Liu}
\affiliation{College of Science, Southern University of Science and Technology, Shenzhen, 518055, Guangdong, China}
\affiliation{China Institute of Atomic Energy, P. O. Box 27547, Beijing 102413, China}

\author{A.~Muñoz}	
\affiliation{%
IGFAE, Universidade de Santiago de Compostela, E-15782, Santiago de Compostela, Spain}

\author{F.~Ndayisabye}
\affiliation{%
FRIB/NSCL Laboratory, Michigan State University, East Lansing, Michigan 48824, USA}

\author{J.~Pereira}	
\affiliation{%
FRIB/NSCL Laboratory, Michigan State University, East Lansing, Michigan 48824, USA}

\author{N.~Rijal}
\affiliation{%
FRIB/NSCL Laboratory, Michigan State University, East Lansing, Michigan 48824, USA}

\author{C.~Santamaria}
\affiliation{%
FRIB/NSCL Laboratory, Michigan State University, East Lansing, Michigan 48824, USA}

\author{D.~Schumann}	
\affiliation{Laboratory of Radiochemistry, Paul Scherrer Institute, Villigen, Switzerland}

\author{N.~Watwood}	
\affiliation{%
FRIB/NSCL Laboratory, Michigan State University, East Lansing, Michigan 48824, USA}

\author{G.~Votta}	
\affiliation{%
FRIB/NSCL Laboratory, Michigan State University, East Lansing, Michigan 48824, USA}

\author{P.~Yin}
\affiliation{College of Physics and Engineering, Henan University of Science and Technology, Luoyang 471023, China}

\author{C.~X.~Yuan}
\affiliation{%
Sino-French Institute of Nuclear Engineering and Technology, Sun Yat-Sen University, Zhuhai 519082, China
}%

\author{Y.~N.~Zhang}
\affiliation{%
Sino-French Institute of Nuclear Engineering and Technology, Sun Yat-Sen University, Zhuhai 519082, China
}%

\date{\today}

\begin{abstract}

Isoscalar dipole transitions are a distinctive fingerprint of cluster structures. A $1^-$ resonance at 7.27(10) MeV, located just below the $\alpha$-emission threshold, has been observed in the deuteron inelastic scattering reactions off $^{10}$Be. The deformation lengths of the excited states in $^{10}$Be below 9 MeV have been inferred from the differential cross sections using coupled channel calculations.  This observed $1^-$ resonance has isoscalar characteristics and exhausts approximately $5\%$-$15\%$ of the isoscalar dipole energy-weighted sum rule, providing evidence for pronounced $\alpha$ cluster structure in $^{10}$Be. The Gamow coupled channel approach supports this interpretation and suggests the near-threshold effect might be playing an important role in this excitation energy domain. The $\alpha+\alpha+n+n$ four-body calculation reproduces the observed enhanced dipole strength, implying that the four-body cluster structure is essential to describe the $1^-$ states in $^{10}$Be.
\end{abstract}

\maketitle

\textit{Introduction.-} 
The clustering phenomenon has attracted much attention because it can appear at different scales of matter, ranging from stars in the universe to systems of microscopic particles~\cite{Genzel, Freer}. 
In atomic nuclei, $\alpha$ clustering is a well-established phenomenon ~\cite{Freer2007, Freer} and can lead to isospin asymmetry, which manifests itself as dipole excitations in nuclei~\cite{Aumann, Iachello}. Such dipole transitions are fingerprints of asymmetrical cluster structures that are commonly found in  nuclei~\cite{Spieker, Chiba}, ranging from light (for example, $^9$Be, $^{10}$Be, $^{12}$C, $^{16}$O)~\cite{Kanada, Kanada2017, Kanada2018, Shikata, Kanada-AMD}, medium~\cite{Tanaka} and heavy mass nuclei~\cite{Daley, Gai}.

Over the last decades, there has been a growing recognition that diverse cluster configurations can arise in asymmetric nuclei. For instance, nuclear systems with excess neutrons can mix different cluster excitation modes due to the various ways that valence neutrons are coupled to the core\cite{Liu, Yang}. With the assistance of theoretical calculations, cluster configurations can be extracted by a careful analysis of their respective dipole transition strength distributions~\cite{11B, Kimura}. 

Low-energy dipole transitions can be excited with sizable cross sections via inelastic scattering reactions and are typically found in the vicinity of the particle-emission thresholds, where the continuum coupling plays an important role and may affect the cluster structure. 
%Therefore, intriguing cluster phenomena can be observed near the particle emission threshold. 
For example, aligned eigenstates have been associated with the threshold effect, where a relatively stable structure is formed, carrying the characteristics of the nearby decay channel. Evidence of this ``alignment'' effect includes the 7.654-MeV Hoyle state in $^{12}$C~\cite{Freer}, the 11.425-MeV state in $^{11}$B~\cite{Ayyad}, and the 3.2-MeV $0^-$ state in $^{12}$Be~\cite{Chen}. 
Novel theoretical approaches, such as the shell model embedded in the continuum (SMEC)~\cite{Bennaceur2000,Okolowicz2003}, Gamow shell model (GSM)~\cite{Michel2002,Michel2021}, and Gamow coupled channel (GCC)~\cite{Wang2021,Wang2023} have been successfully applied to these threshold-aligned states.

Neutron-rich Be isotopes provide an ideal playground for studying the interplay between cluster structure and neutron correlations because of the presence of valence neutrons outside a $2\alpha$ (or $^8$Be) core. $^{10}$Be nucleus can be simply considered as a combination of two valence neutrons coupled to the $2\alpha$ core. The ground-state (g.s.) molecular structure has been experimentally validated~\cite{Li}. Several low-lying dipole excitations in $^{10}$Be have been predicted~\cite{Kanada, Shikata}. However, only one low-lying $1^-_1$ state at 5.960 MeV has been experimentally observed and has a very small B(E1) value ~\cite{nndc, Mattoon, Jiang}. 

The low-lying structure of $^{10}$Be has been observed up to 9 MeV with the missing mass method via inelastic scattering of deuterons thanks to the large acceptance for the charged particles of the Active-Target Time Projection Chamber (AT-TPC)~\cite{Bradt} coupled to SOLARIS~\cite{SOLARIS}. A dipole resonance with isoscalar characteristics has been observed at 7.27(10) MeV in $^{10}$Be, which is located just below the $\alpha$ emission threshold ($S_{\alpha}=7.41$ MeV) and likely corresponds to the known 7.37-MeV state observed before. This $1_2^-$ state has an enhanced dipole strength and exhausts 5$\%$-15$\%$ of the isoscalar dipole energy-weighted sum rule (IS-EWSR). The observed enhanced isoscalar dipole (ISD) strength in this resonance validates its pronounced cluster structure and provides new evidence for the threshold ``alignment'' effect. The structure of this dipole resonance seems to be affected by the $\alpha$-, $1n$-, and $2n$-emission thresholds, resulting in the mixing of the different cluster configurations. The $\alpha+\alpha+n+n$ four-body cluster structure and different neutron correlation configurations are essential to describe it.

\textit{Experiment.-} The experiment was carried out at the ReA6 reaccelerator beam facility of the National Superconducting Cyclotron Laboratory (NSCL). The scattered deuterons were measured using the AT-TPC~\cite{Bradt}. A long-lived $^{10}$Be beam at 9.6 MeV/u was produced with an intensity of approximately 2000 particles per second (pps) and a purity close to $100\%$. The beam intensity and purity were monitored using an ionization chamber positioned upstream of the AT-TPC, filled with CF$_4$ gas at 200 Torr. %, and has a length of 35 mm between two layers of PPTA plastic windows of  12 $\mu$g/cm$^2$. This beam enters the AT-TPC through a 1-cm-diameter 12 $\mu$g/cm$^2$ poly-phenylene terephthalamide (PPTA) window. 
The active volume of the AT-TPC was filled with 600 Torr of pure deuterium gas, which has an areal density of $10^{20}$/cm$^2$ and equivalent to a 35-mg/cm$^2$ conventional solid (CD$_2)_n$ target. The $^{10}$Be beam energy at the center of the detector was 9.06 MeV/u, and the beam energy variation in the detector was approximately 10 MeV, which was corrected event-by-event according to the vertex position. The detector was installed inside the SOLARIS solenoidal spectrometer which provides a magnetic field of 3.0 T. The cathode of the AT-TPC is defined as $z=0$ on the beam axis. The pad plane was located at $z=100$ cm %and consists of 10240 triangular pads, which 
provided track information on the plane perpendicular to the beam direction. The signal from each pad was sampled at a frequency of 3.125 MHz and was used to deduce the track information on the beam axis. %The electron drift velocity was measured to be about 0.9 cm/$\mu$s, so each sample corresponds to 0.29 mm in the beam direction.  

\begin{figure}
\includegraphics[width=1.0\columnwidth]{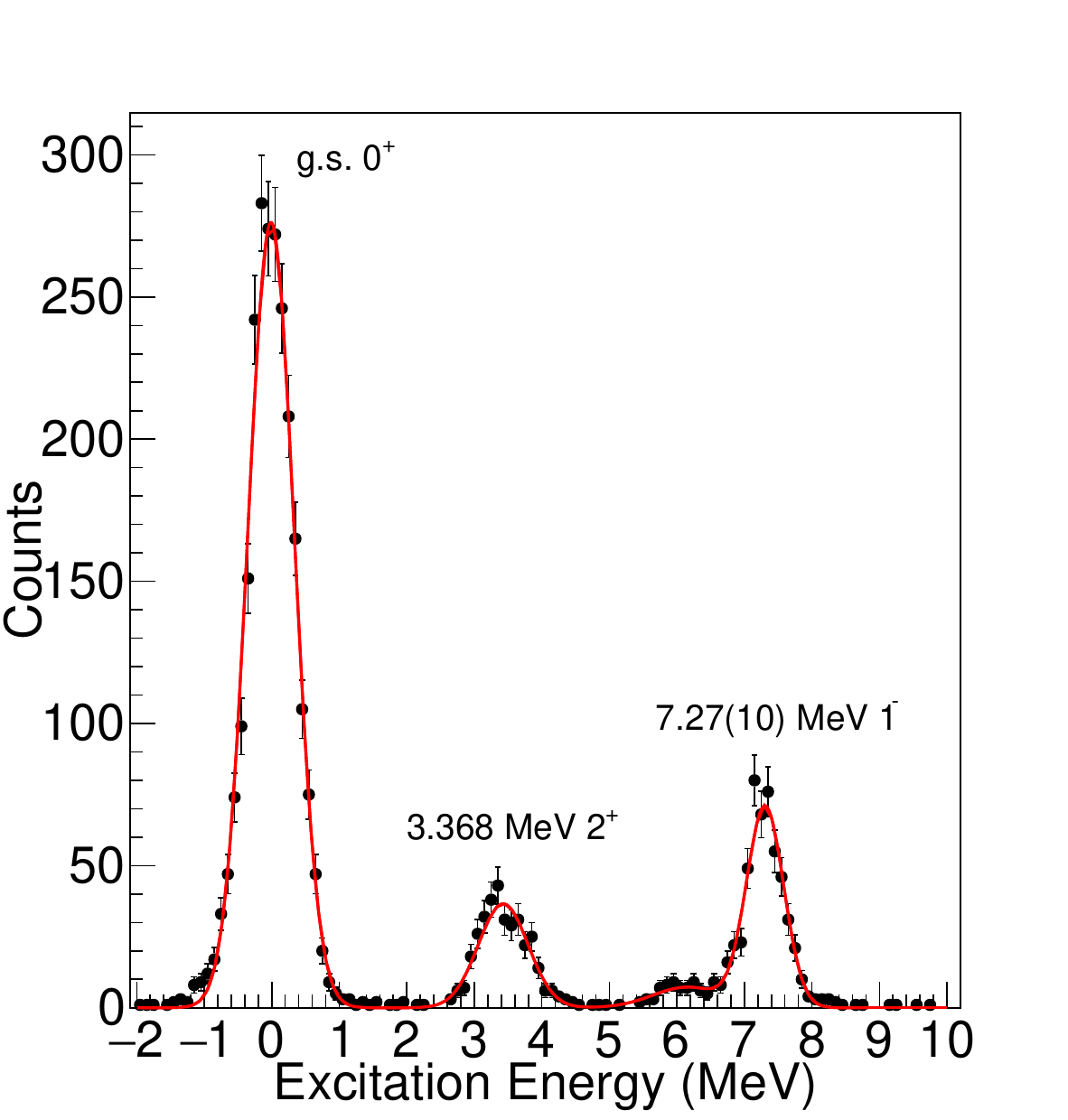}
\caption{\label{fig:kin} The excitation spectrum of $^{10}$Be populated via the $^{10}$Be scattering on deuterons. The peaks are labeled by their excitation energies (in MeV) and their spin-parity assignments. The spectrum was gated on c.m. angles $20^{\circ}$ to $80^{\circ}$, corresponding to a maximum of 15 MeV for the elastic channel.}
\end{figure}

The signal from the pad plane was amplified, sampled, and read out by a dedicated electronic system~\cite{GET} triggered by the micromegas mesh signal. A 2-cm-radius region of the pad plane illuminated by the beam particles was set to less than $10\%$ amplification factor compared to other pads, so the beam events do not trigger or saturate the system. The beam rate was monitored throughout the experiment by recording the downscaled beam signal triggered by the ionization chamber. 

%The AT-TPC coupled with SOLARIS was used to record deuterons from 
Kinematics of deuterons scattered off $^{10}$Be were inferred from the point cloud using an algorithm based on a linear quadratic estimator (Kalman filter), achieving an average efficiency of approximately 70$\%$ in the 7-10~MeV excitation energy range. The details of this algorithm can be found in Ref.~\cite{Ayyad_tracking, Ayyad_ATTPC}.  based on their $B\rho$ values versus their energy losses. Protons, deuterons, and tritions are clearly isolated based on their $B\rho$ values versus their energy losses, where deuterons from the elastic/inelastic scattering reactions were selected. The energy versus the angles of the selected events can be found in Ref.~\cite{Ayyad_tracking}. %The last 20 cm of the detector was not included in the analysis because of the decreasing acceptance. 

The spectrum resulting from the elastic and inelastic scattering of $^{10}$Be on deuterons was obtained based on reaction kinematics, as shown in Fig. ~\ref{fig:kin}, which has a resolution of approximately 700 keV (FWHM). The first three peaks in the spectrum correspond to the ground state, and excited states at 3.368 MeV ($2_1^+$) and the doublets 5.958 MeV ($2_2^+$)/5.960 MeV ($1_1^-$), respectively. The fourth peak at 7.27(10) MeV may correspond to the 7.37-MeV state observed before.  Particle energy below 15 MeV was calibrated by the known excitation energies and also verified by the data of the $^{10}$Be$(d,p)$ reaction channel.

\begin{figure}
 \includegraphics[width=1.0\columnwidth]{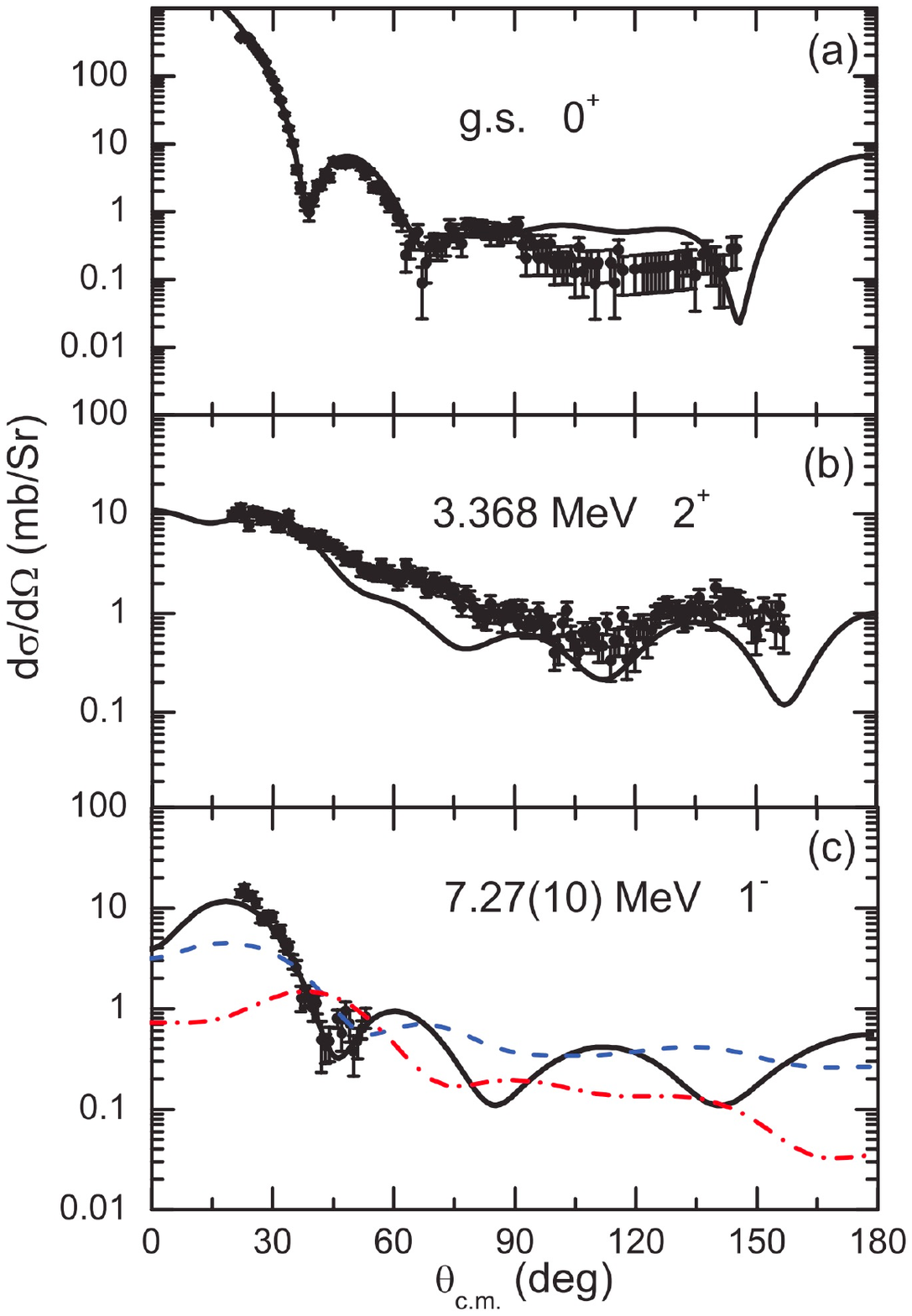}
\caption{\label{fig:cc}Coupled channel calculations compared to the data for (a) 9.06 MeV/u $^{10}$Be + $d$ elastic scattering, (b) $^{10}$Be + $d$ inelastic scattering to the 3.37-MeV $2^+$ level of $^{10}$Be, and (c) $^{10}$Be + $d$ inelastic scattering to the 7.27-MeV level of $^{10}$Be, considered as a dipole. Quadrupole ($2^+$, blue dashed line) and octupole ($3^-$, red dotted-dashed line) angular distributions for the 7.27-MeV state are also compared.
}  
\end{figure}
\textit{Identification of the dipole resonance.-}
The absolute differential cross sections were extracted from the present data, as shown in Fig.\ \ref{fig:cc}. Owing to the nearly $4\pi$ solid angle of the AT-TPC, we obtained the angular distributions for elastic scattering and inelastic scattering to the 3.37-MeV state over an angular range of $20^{\circ}<\theta_{cm}<120^{\circ}$. For the 7.27-MeV state, the data above $\theta_{cm}=50^{\circ}$ are not presented because of the prominent background. Each $1^{\circ}$ was divided into one angular bin, owing to the outstanding angular resolution of the detector. The cross sections below $\theta_{cm}<20^{\circ}$ are not shown in Fig.\ \ref{fig:cc}, because of the decreasing acceptance of the low-energy deuterons. The uncertainties of the absolute cross sections are dominated by the systematic uncertainties resulting from the beam intensity and the acceptance of the detector. The former should be within $10\%$ owing to the good efficiency of the ionization chamber. The latter was determined to be less than 5$\%$ by a simulation. %that accounted for the staggered division of charge deposited on adjacent triangular pads, energy-dependent energy loss, lateral diffusion of drift electrons, and digitization process. 

The measured angular distributions were analyzed within the coupled channel (CC) formalism using the code {\sc fresco}~\cite{Tho88}. 
Initial calculations only included coupling to the 3.37-MeV $2^+$ level of $^{10}$Be. The strength of Coulomb coupling was fixed using the adopted B(E2) value of Ref.~\cite{Ram01}. The nuclear deformation length ($\delta_2$) and the diagonal optical model potential (OMP), were adjusted to obtain the best overall fit to both the elastic and inelastic scattering angular distributions using the {\sc sfresco} package.

Coupling to the 7.27-MeV level of $^{10}$Be was then added, and initial attempts to fit the angular distribution for inelastic scattering to this state considered the spin-parity assignments of either $J^\pi = 2^+$ or $3^-$.  %that is, \ quadrupole or octupole excitation from the ground state of $^{10}$Be. 
However, neither of these resulted in a satisfactory description. Therefore, a fit assuming a $1^-$ assignment (dipole excitation) was performed (see Fig.~2c). %Because the dipole nuclear form factor is not of the standard form (see Ref.~\cite{Sat83}), it was calculated externally using {\sc fresco} code. 
The form factor was obtained by deforming the diagonal OMPs using the method described in Ref.~\cite{Har81}. 
An initial estimate of the dipole deformation was calculated using Eq. \ (5) of Ref.~\cite{Har81}. Together with the best-fit diagonal potential obtained above, this was used to calculate the dipole coupling potential according to Eq. \ (6) of Ref.~\cite{Har81}. The dipole deformation length was adjusted to provide the best description of the inelastic scattering data for excitation at the 7.27-MeV level, and the parameters of the diagonal potential were searched to optimize the fit to the entire data set.  An acceptable description of the whole data set was obtained with a set of consistent parameters.

The final results of this procedure are compared with the elastic and inelastic scattering data in Fig.\ \ref{fig:cc}. The corresponding nuclear deformation lengths are $\delta_2=1.89$ fm for coupling to the 3.37-MeV $2^+$ level in agreement with the value in the literature $\delta_p\sim1.89(18)$ fm~\cite{Iwasaki}, obtained from proton inelastic scattering. $\delta_1 = 0.76$ fm was deduced for coupling to the 7.27(10)-MeV level. The uncertainties in both values are about $25$\%, dominated by the variation of the OMPs and the uncertainties of the absolute cross sections. The value of $\delta_1$ exhausts approximately $5\%$-$15\%$ of the IS-EWSR~ \cite{Har81}. Based on these results, an unambiguous assignment of $1^-$ spin-parity to the observed 7.27(10)-MeV level in $^{10}$Be was made. 

It worth noting that the assignment of the dipole resonance was driven by the slope of the downward trend of the first maximum ($20^{\circ}<\theta_{cm}<50^{\circ}$). At these angles, the shape of the angular distributions is not very sensitive to the OMPs, while the impact is significant at larger c.m. angles. Attempts have been made with different OMPs, but same conclusion was drawn~\cite{An, Han}.

There are four states at about 6.0 MeV, including the $2_2^+$ (5.958 Mev), $1_1^-$ (5.960 Mev), $0_2^+$ (6.179 Mev), and $2_1^-$ (6.263 MeV) states, which could not be resolved with present resolution. However, according to Ref.~\cite{Kanada2020}, the $1_1^-$ state should have much smaller cross sections than the $2_2^+$ state, and the centroid of the peak located at 6.0 MeV also rules out the strong population of the $0_2^+$ and $2^-_2$ states. As expected, its angular distribution agrees reasonably well with the calculation assuming a quadrupole transition.

\begin{figure}[htb]
  \includegraphics[width=0.7\columnwidth,angle=270]{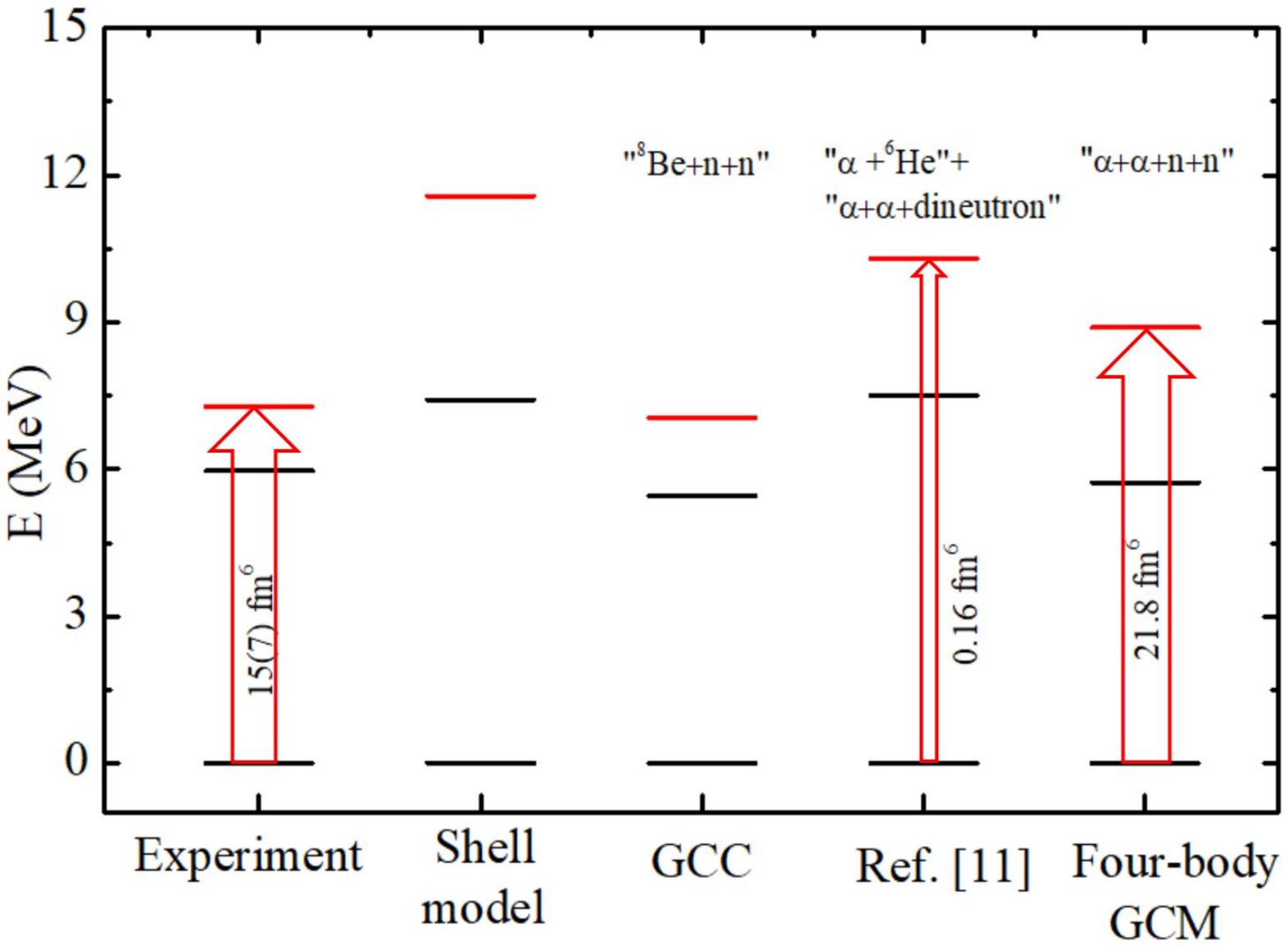}
  \caption{The predicted excitation energies of the $^{10}$Be dipole states and corresponding ISD strength (red arrow) by different theories compared to the experimental values. See text for more details. }
  \label{GCC}
\end{figure}

\textit{Theoretical calculations and discussions.-}

The 7.27-MeV state has been observed as the second $1^-$ state of $^{10}$Be and its width was determined to be less than hundreds of keV. One $3^-$ state was previously observed at 7.37 MeV, which like corresponds to the presently observed state at 7.27(10)~MeV. It was noted that the $1^-$ spin-parity of the 7.37-MeV state is inline with all existing measurement, except that the spin is conflict with the neutron resonance scattering reaction in Ref.~\cite{Lane}.  The present result cast doubt on the previous spin assignment. Event if we consider the observed resonance as a new one, the population upper limit of the 7.37-MeV $3^-$ and 7.54-MeV $2^+$ states were determined to be about 3$\%$ and $5\%$, respectively, from the angular distribution and by fitting the peak. It is worth mentioning that this new assignment does not reconcile with the spin-parity of the correspondent isobaric analog states 3$^{-}$/2$^{+}$ of the 8.887/8.895 MeV states in $^{10}$B. However, both states have a pronounced $\alpha$ width, a strong evidence of clustering~\cite{Kuchera2011}.

The effective interaction shell model calculation using the YSOX interaction~\cite{YSOX} has been performed to study the low-lying $1^-$ states of $^{10}$Be. The calculated excitation energies are plotted in Fig.~\ref{GCC}. The second $1^-$ state obtained in the shell model calculation is a little higher than the newly reported experimental result. The B(E1) value of the first $1^-$ state was predicted to be 0.013 e$^2$fm$^2$, and is much higher than the experimental value B(E1)= $1.6\times 10^{-6}\sim 3\times10^{-4}$ e$^2$fm$^2$ obtained in Ref.~\cite{Mattoon}. No information on the ISD strength can be obtained.  This discrepancy is possible because the $\alpha$ cluster structure was not incorporated in these calculations. 

The observed 7.27(10)-MeV resonance is located close to the $S_n$, $S_{\alpha}$, and $S_{2n}$ of $^{10}$Be at 6.812, 7.409, and 8.476 MeV, respectively. Therefore, mixing of different cluster configurations may occur because of the threshold effect, including the two-body $\alpha+^6$He, three-body $^8$Be+n+n, or even four-body $\alpha+\alpha+n+n$ cluster structures. To simultaneously incorporate the continuum coupling effect and the $2\alpha$ cluster structure, we have performed calculations within the GCC approach assuming a system of a deformed $^8$Be (2$\alpha$) core plus two neutrons. It is constructed in Jacobi coordinates with a Berggren basis, incorporating the three-body feature and continuum effect~\cite{Wang2018,Pfutzner2023,Zhou2022}. The ground and $2^+$ states of  $^8$Be are coupled to the valence neutrons through a non-adiabatic rotational coupling~\cite{Wang2018, Wang2022}. %
The excitation energies of the first and second $1^-$ states (Fig.~3) agree with the experiment, supporting that the continuum affects the excitation energy substantially and is essential here. The general feature of $^{10}$Be is described reasonably well by assuming a $2\alpha$ cluster plus two valence neutrons. However, since the motion of the 2$\alpha$ against each other is fixed in this calculation, the GCC approach cannot give a reliable dipole strength. 

\begin{figure}[htb]
   \includegraphics[width=0.9\columnwidth]{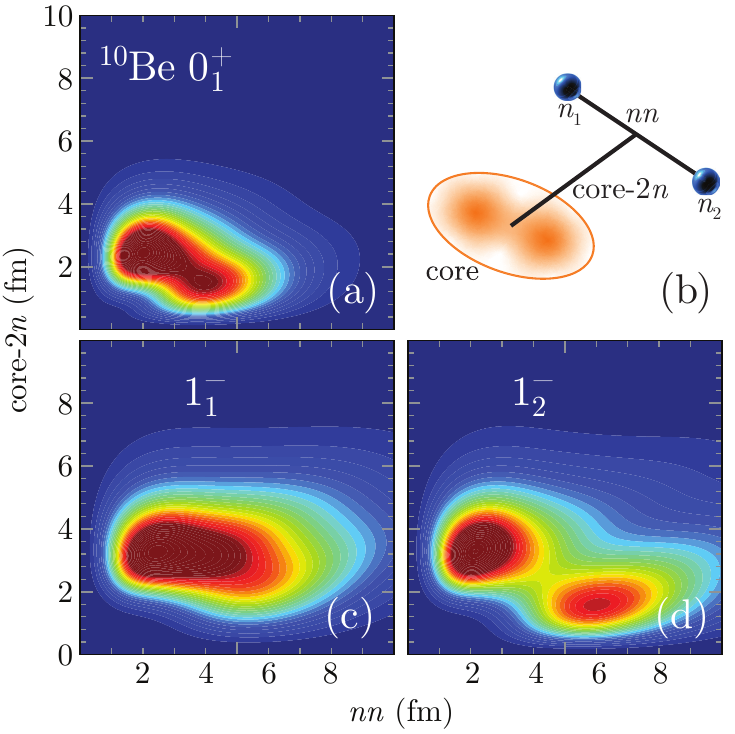}
  \caption{Two-nucleon density distributions (in fm$^{-2}$) in Jacobi coordinates predicted by GCC for the g.s. and low-lying $1^-$ states in $^{10}$Be. }
  \label{GCC2}
\end{figure}

Fig.~4 shows the Jacobi-coordinate density distribution of the g.s. (0$^{+}_1$),  and the two $1^{-}$ states in $^{10}$Be. It can be seen that the first and the second $1^-$ are showing very different features. 
The two neutrons of the first $1^-$ state are less correlated due to the more uniform distribution. Still, the second $1^-$ state shows features similar to the $0^+$ states demonstrating a strong dineutron correlation. This also indicates rapid structure changes when crossing the different thresholds.  

Recently, the low-lying $1^-$ states were investigated utilizing the microscopic cluster model, which incorporates the ($\alpha+^6$He and $\alpha+\alpha$+dineutron) configurations. Specifically, the focus was on studying the dipole excitations in the three-body cluster dynamics and the results are plotted in Fig.~3~\cite{Kanada-AMD}. 
However, the predicted E1 strength (0.023 e$^2$fm$^2$) was two orders of magnitude higher than the experimental value, and the dipole strength (0.16 fm$^6$) of the $1_2^-$ state is underestimated. This is probably because the two valence neutrons are treated as one cluster but the degree of freedom of the two neutrons was not included, which may conversely affect the two $\alpha$ motions. Indeed, the different two-neutron configurations are also noted in the result of the GCC approach, as stated above.

To encompass the dynamics of the (2$\alpha$) cluster structure along with the motions of the two valence neutrons, the four-body cluster model, ($\alpha+\alpha+n+n$), was implemented utilizing GCM Brink wave functions. It is noteworthy that the interaction parameters adopted in this study are in accordance with those specified in Ref.~\cite{Kanada-AMD}.
An intriguing aspect is that, in the description of the first two $1^-$ states, the new four-body calculations exhibit significant discrepancies if compared to the calculations assuming a dineutron configuration reported in Ref.~\cite{Kanada-AMD}.

The excitation energies and transition probabilities with the four-body calculation are shown in Fig.~3.  The predicted B(E1) value of $1^-_1$ state ($1.56\times 10^{-5}$e$^2$fm$^2$) has a similar amplitude to the experimental value. The predicted dipole strength is also consistent with our experimental observations. The improvement compared to the tri-cluster model indicates that the four-body dynamic and the two neutrons' degree of freedom are essential to describe these dipole excitations.  
The ISD strength in the $1^-_2$ state is an indication of the enhanced cluster structure of $^{10}$Be.  However, the predicted dipole strength in the $1_1^-$ state ($\sim$21 fm$^6$) was not observed experimentally, which requires further understanding. 
Development of the four-body calculation incorporating the continuum coupling effect is still required. This observed 7.27 MeV $1_2^-$ dipole resonance not only retains the prominent clustering ground state structure~\cite{Li}, but also shows a configuration mixing of the two-body $^6$He$+\alpha$, three-body $^8$Be+n+n, and four-body $\alpha+\alpha+n+n$ cluster structures, which emerge possibly due to the near-threshold effect of the $1n$-, $2n$- and $\alpha$-emission thresholds.

\textit{Summary.-} 
In summary, the deuteron inelastic scattering of $^{10}$Be has been measured with the AT-TPC coupled with SOLARIS at the ReA6 beamline of NSCL. A $1^-$ resonance with enhanced isoscalar dipole strength has been observed at 7.27(10) MeV in $^{10}$Be, which is located just below the $\alpha$-emission threshold and likely corresponds to the 7.37 MeV observed previously. It exhausts about 5$\%$-15$\%$ of the isoscalar dipole strength, providing evidence for the pronounced $\alpha$-cluster structure of $^{10}$Be and for the threshold alignment effect. The GCC approach reproduces the excitation energies of the dipole states reasonably well, and the observed enhanced dipole strength was reproduced by the $\alpha+\alpha+n+n$ four-body calculation, indicating that four-body cluster assumption and the continuum coupling effect is essential in the description of $^{10}$Be nucleus. 
The AT-TPC coupling with SOLARIS provides a powerful experimental tool for future measurements of direct reactions with rare exotic beams to enhance our understanding of this phenomenon.

\textit{Acknowledgement.-} 
The authors would like to acknowledge the operation staff at ReA6 (NSCL) for providing the beam. This material is based upon work supported by National Superconducting Cyclotron Laboratory, which has been a major facility fully funded by the National Science Foundation under award PHY-1565546; the U.S.\ Department of Energy, Office of Science, Office of Nuclear Physics, under Contract Number DE-AC02-06CH11357 (Argonne), DE-SC0020451 (FRIB); the Spanish Ministerio de Economía y Competitividad through the Programmes “Ramón y Cajal” with the grant number RYC2019-028438-I; the National Key Research and Development Project under grant No. 2022YFA160230; Grant RYC2020-030669 funded by MCIN/AEI/ 10.13039/501100011033; Grant PID2022-142557NA-I00 funded by MCIN/AEI/ 10.13039/501100011033. This work was Supported by National Natural Science Foundation of China under Contract No.\,12147101, 12475120, 12435010.  SOLARIS is funded by the DOE Office of Science under the FRIB Cooperative Agreement DE-SC0000661. This material is based upon work supported by the National Key Research and Development Program (MOST 2022YFA1602303).

\def\bibindent{0em}

%\begin{appendices}
\hspace{1cm}

{\bf Appendix}

In the CC calculations, the coupling parameters for the real and imaginary parts of the transition potential, $\beta_\mathrm{R}$ and $\beta_\mathrm{I}$, respectively, were related such that $\beta_\mathrm{R} R_\mathrm{R} = \beta_\mathrm{I} R_\mathrm{I}$, i.e.\ the real and imaginary deformation lengths were equal. The fitted optical parameters used in the calculation are listed in Table I.

\begin{table*}
\centering
\caption{\label{tab:omp} Parameters of the best-fit optical model potential. Notation follows that of Ref.\ \cite{Han}.
All radii are defined as: $R_\mathrm{i} = r_\mathrm{i} \; 10^{1/3}$. } 
\begin{tabular}{cccccccccc}
\hline
$V$ [MeV] &$ r_\mathrm{0}$ [fm] & $a_\mathrm{0}$ [fm] & $W$ [MeV] & $r_\mathrm{W}$ [fm] & $a_\mathrm{W}$ [fm] & $W_\mathrm{D}$ [MeV] & $r_\mathrm{D}$ [fm] & $a_\mathrm{D}$ [fm] & $r_\mathrm{C}$ [fm] \\
\noalign{\smallskip}\hline\noalign{\smallskip}
122.2 & 1.244 & 0.450 & 14.42 & 1.366 & 0.300 & 2.30 & 2.501 & 0.524 & 1.30 \\
\hline
\end{tabular}
\end{table*}

In the GCC approach, the quadrupole deformation parameter $\beta_2$ is chosen as 0.9. For the nuclear two-body interaction between valence neutrons we took the finite-range Minnesota force with the original parameters \cite{Thompson1977}. The effective core-valence potential has been taken in a WS form including the spherical spin-orbit term with ``universal'' parameter set \cite{Cwiok1987}. The depth of the WS potential has been readjusted to the experimental value of two-neutron separation energy. The calculations have been carried out in the model space of $\max(\ell_x,\ell_y)\leq 7 $ with the maximal hyperspherical quantum number $K_{\max} = 16$, where $\ell_x$ $\ell_y$ are the orbital angular momenta in Jacobi coordinates \cite{Wang2017}. To investigate resonances in the GCC framework, we used the Berggren basis for the $K_{\max} \leq 6$ channels and the harmonic oscillator basis with the oscillator length $b =$ 1.75 fm and $N_{\max} =$ 40 for the higher-angular-momentum channels. The complex-momentum contour  for the Berggren basis is defined as $k=0 \rightarrow 0.3-0.09i \rightarrow 0.5-0.05 \rightarrow 0.8 \rightarrow 8 $ (all in fm$^{-1}$) and discretized with 30 scattering states for each segment. 
%The excitation energy of each state from this GCC calculation is shown in Fig.~\ref{GCC}.
%\end{appendices}
\end{document}